\documentclass[final,3p,times,twocolumn]{elsarticle}
 \biboptions{comma,sort&compress}

\usepackage{graphicx}
\usepackage{here}

\def\beq{\begin{equation}}
\def\eeq{\end{equation}}
\def\bea{\begin{eqnarray}}
\def\eea{\end{eqnarray}}
\def\bq{\begin{quote}}
\def\eq{\end{quote}}

\def\nnb{\nonumber}

\def\nnb{\nonumber}
\def\la{\langle}
\def\ra{\rangle}
\def\nin{\noindent}
\def\ba{\vspace*{-0.2cm}\begin{array}}
\def\ea{\end{array}\vspace*{-0.2cm}}

\def\b{$\bullet~$}
\def\als{\alpha_s}

\def\gg2{ \la\alpha_s G^2 \ra}
\def\gg3{g^3f_{abc}\la G^aG^bG^c \ra}
\def\ggg4{\la\als^2G^4\ra}

\journal{Nuc. Phys. (Proc. Suppl.)}

\begin{document}

\begin{frontmatter}

\title{Mini-review on QCD spectral sum rules$^*$} 

 \author[label1]{Stephan Narison
 }
   \address[label1]{Laboratoire
Univers et Particules, CNRS-IN2P3,  
Case 070, Place Eug\`ene
Bataillon, 34095 - Montpellier Cedex 05, France.}
\cortext[cor1]{Talk given at  QCD 14 (29 june - 3 july, Montpellier - France).}
\ead{snarison@yahoo.fr}


\pagestyle{myheadings}
\markright{ }
\begin{abstract}
\noindent
Taking the example of the most popular and well-established  {\it Borel / Laplace / Exponential sum rule (LSR)}, I shortly review 
some of its recent applications in hadron physics  namely  the estimates of non-perturbative condensates, the  determination of the light  and heavy quark masses, the
extraction of the heavy-light decay constants, the estimates of charmonium and bottomium molecule  masses and the properties of scalar gluonium.  In addition to the standard $\tau$ (sum rule variable) and $t_c$ (QCD continuum threshold) stablity criteria,  I introduce  a new stability criterion versus the arbitrary QCD subtraction point $\mu$ for extracting the optimal results when radiative QCD corrections are included.  Future improvements on further uses 
of QCD spectral sum rules (QSSR) are commented.
 \end{abstract}
\begin{keyword}  QCD spectral sum rules, Non-perturbative methods. 


\end{keyword}

\end{frontmatter}
\section{Introduction}
 \nin
In this talk, I review some recent developments of the SVZ sum rules, which I often call 
{\it QCD spectral sum rules (QSSR)} instead of the popular name {\it  QCD sum rules} for two reasons: the word {\it spectral}
is essential as one works in this approach with spectral functions. The 2nd reason is that {\it QSSR} 
is near the abbreviation of the former Soviet Union  state (USSR) from which originally  come the three inventors of this method.
To my opinion, the SVZ sum rule \cite{SVZ} is one of the most important discovery of the 20th century in high energy physics phenomenology (Sakurai price to SVZ in 1999, Pomeranchuck prices in 2010 for V. Zakharov and in 2013 for M. Shifman) with a wide successful spectrum of applications in hadron physics \cite{SHIF,ZAK,DOSCH,SN10A,ZAKA,BERTL,IOFFE,REV,SNB1,REVMOL,CONF,SNMOL14} as it permits to tackle in a semi-analytic way the complicated hadron dynamics using the QCD fundamental parameters despite our ``poor understanding" of QCD confinement. 
The approaches of spectral sum rules are not new. In the year of 60 - 70, there were many different sum rules
based on the old idea of duality \cite{DUALITY}, superconvergence \cite{WEINBERG} and smearing \cite{POGGIO}. 
Some attempts to improve these pre-QCD sum rules using perturbative QCD  have been done \cite{FNR} using finite energy sum rules (FESR) contour techniques \`a la Shankar \cite{SHANKAR} for testing the breaking of the convergence of the Weinberg sum rules  \cite{WEINBERG} when the perturbative current quark masses are switch on. As a consequence, some combinations of superconvergent sum rules for non-zero light quark masses have been  proposed \cite{DUNCAN,FNR}. 
In the same period, SVZ have presented a new sum rule with an exponential weight which permits to optimize the contribution of the low-energy region of the sum rule but has also extended the  QCD results to lower energy by including, via an operator product expansion (OPE), non-perturbative condensates which can mimics in this region the main features of QCD confinement.
\section{QCD spectral sum rules}
 \nin
From these short historical reminders, the SVZ sum rules are, like its predecessors, alternative  improvements of the well-known K\"allen-Lehmann dispersion relation, which for a hadronic two-point correlator  $\Pi(Q^2)$ reads:
\bea
\Pi(Q^2)&\equiv &i\int d^4x~e^{iqx}\la 0| {\cal T} J(x) J^\dagger (0)|0\ra
\nnb\\
&=&\int_{t_<}^{\infty}{dt\over t+Q^2+i\epsilon}{1\over\pi}{\rm Im} \Pi(t)+...
\label{eq:dispersion}
\eea
where $Q^2\equiv -q^2>0$, ... means arbitrary subtraction terms which are (in general) polynomial in $Q^2$; $J(x)$ is any hadronic current with definite quantum numbers built from 
quarks and/or gluon fields; $t_<$ is the hadronic threshold. This well-known dispersion relation is a very important {\it bridge between theory and experiment} as it relates $\Pi(Q^2)$ (which can be calculated in QCD provided that $Q^2$ is much larger than the QCD scale $\Lambda^2$), with its imaginary part ${\rm Im} \Pi(t)$ which can be measured at low energy from experiments. 
At present, there are different variants of QCD spectral sum rules in the literature. They have been mainly inspired  from FESR \cite{DUALITY,FNR,FESR,PEROTTET} and $\tau$-decay semi-hadronic width \cite{BNP}. 
 \vspace*{-0.25cm}
\subsubsection*{\b  Borel / Laplace / Exponential sum rules (LSR)}
\nin
In this
paper, we shall concentrate on the recent and modern applications of the original Borel / Laplace or Exponential most popular SVZ sum rules\,\cite{SVZ}\,\footnote{Its first radiative corrections have been first evaluated in \cite{PSEUDO} where the PT series has the property of an (inverse) Laplace transform. Its  interpretation in quantum mechanics has been discussed in\,\cite{BELL}.}:
\beq
 {\cal L}(\tau)= \int_{t_<}^{\infty} dt~{e}^{-t\tau}~{1\over\pi}{\rm Im} \Pi(t)~.
\label{eq:lapl}
\vspace*{-0.25cm}
\eeq
This sum rule is obtained  when one  takes an infinite number of derivatives $n$ of the correlator in $Q^2$ but keeping the ratio $Q^2/n\equiv M^2\equiv \tau^{-1}$ (sum rule variable) fixed in the dispersion relation in Eq. (\ref{eq:dispersion}). In this way, one can eliminate the substraction terms (polynomial in $Q^2$) and the dispersion becomes an exponential sum rule which has the nice feature to enhance the lowest resonance contribution and to suppress the ones of higher hadron masses (similar exponential suppression is used on the lattice for extracting the lowest meson parameters).
 \vspace*{-0.25cm}
\subsubsection*{\b  Ratios and Double Ratios of sum rules (DRSR)}
\nin
Another related sum rules are the ratio and double ratios of sum rules:
\beq
 {\cal R}(\tau)\equiv -{d\over d\tau}\log {\cal L}(\tau)~~{\rm and}~~ r_{ij}={{\cal R}_i\over {\cal R}_j}~,
 \label{eq:ratio}
 \eeq
for two different channels $i$ and $j$.  The former is often used in the literature for extracting the lowest ground state hadron mass as  at the optimization point (discussed later on) $ {\cal R}(\tau_0)\simeq M_R^2$ where we have a cancellation of the leading $\alpha_s$ PT corrections, 
while the second is useful for extracting with a good accuracy the splittings of hadron masses
due, in addition, to the cancellation of the leading non-flavoured terms in the QCD expression of the sum rule~\cite{SNDRSR}.
 \section{The original SVZ - Expansion}
 \nin
The SVZ ideas are not only related to the discovery of sum rules.  More important,  they have proposed a new way to phenomenologically parametrize (approximately) the non-perturbative (confinement) aspect of QCD beyond perturbation theory using an operator product expansion (OPE) \`a la Wilson in terms of the vacuum condensates of quark or/and gluons of higher and higher dimensions. In this way, the two-point correlator can be parametrized as:
\beq
 \Pi(Q^2)
 =\sum_{d=0,2,4,...} {C_{d} \la 0| O_{d} |0\ra \over Q^{d}}~,
\eeq
where $d$ is the dimension of the non-perturbative condensates $ \la 0| O_{d} |0\ra$ and $C_d$ the corresponding Wilson coefficients calculable using perturbation theory (PT). They are classified in Table \ref{tab:param} with their fitted values from hadronic data. 
In addition to the well-known quark condensate $\la\bar\psi\psi\ra$ entering to the Gell-Mann-Oakes-Renner relation: 
$$
(m_u+m_d)\la \bar uu+\bar dd\ra=-m_\pi^2f_\pi^2:~~~~ f_\pi=(130.4\pm 0.2)~{\rm MeV},
$$
SVZ\,\cite{SVZ,ZAK}  have advocated the non-zero value of the gluon condensate $\la\alpha_s G^2\ra$ (which they have found to be 0.04 GeV$^4$ from charmonium data) and of some other higher dimension condensates. 
\vspace*{-0.5cm}
{\scriptsize
\begin{center}
\begin{table}[hbt]
\setlength{\tabcolsep}{0.2pc}
\begin{tabular}{llll}
\hline 
\footnotesize  Condensates &\footnotesize $d$&\footnotesize Values [GeV]$^{   d}$&\footnotesize Sources\\
\hline
   $\la\bar uu\ra(2)$&\footnotesize 3&\footnotesize$-(0.276\pm 0.07)^3$&\footnotesize  $\pi$ \cite{SNL14,SNB1},
\\
$\la \bar dd\ra/\la\bar uu\ra$&\footnotesize 0&\footnotesize$1-9\times 10^{-3}$&\footnotesize $\pi$ \cite{SNB1}, \\
$\la \bar ss\ra/\la\bar dd\ra$&\footnotesize 0&\footnotesize$0.74(3)$&\footnotesize $K$ \cite{SNB1} $\oplus$ light \cite{DOSCH}\\
&&&\footnotesize    and heavy \cite{HBARYON} baryons\\
 $\la\alpha_s G^2\ra$&\footnotesize 4&\footnotesize$7(2)10^{-2}$ &\footnotesize   $e^+e^-$\,\cite{EIDELMAN,LNT,SNI} $\oplus$\\
 &&&\footnotesize$\Upsilon-\eta_b$\,\cite{SNHeavy},~$J/\psi$ \cite{SNH12,YND}\\
$g\la \bar\psi G\psi\ra $&\footnotesize 5&\footnotesize$M^2_0=0.80(2)$ &\footnotesize   light baryons\,\cite{IOFFEBAR,DOSCH}, $B,B^*$\,\cite{SNBST88}\\
$g^3\la G^3\ra$&\footnotesize 6&\footnotesize$(8.2\pm 2.0)\la \alpha_s G^2\ra$&\footnotesize    $J/\psi$ \cite{SNH12}
 \\
$\rho\alpha_s\la\bar\psi\psi\ra^2$&\footnotesize 6&\footnotesize$(4.5\pm 0.3)10^{-4}$ &\footnotesize  $e^+e^-$\,\cite{PEROTTET,LNT,SNI},~$\tau$-decay \cite{SNTAU}\\
\hline
\end{tabular}
 \caption{\scriptsize  Values of the QCD condensates from QSSR: $g\la \bar\psi G\psi\ra\equiv M_0^2\la\bar \psi\psi\ra$; $\rho$ indicates the deviation from vacuum saturation. }
 \label{tab:param}
\end{table}
\end{center}
 }
\vspace*{-1cm}
\nin
\section{Tachyonic gluon mass beyond the SVZ expansion}
\nin
In a series of papers, Zakharov et al \cite{ZAK,CNZ,SN05,SNZ0,SNZ} have  
parametrized the contribution of the non-calculated
large order terms of the PT series by the introduction of a quadratic term associated to the tachyonic gluon
mass. The presence of this ``operator" is a priori  shocking as a such term is forbidden by gauge invariance in QCD. However,  this term is not of infrared origin like some other non-perturbative condensates but {\it it is dual to the sum of higher order  ultra-violet terms of the PT series}.  Better the series is known
, lesser is the strength
of this term which can vanishes after some high order terms of the PT series \cite{LEE}.  Its presence 
is seen in different dual AdS/QCD models \cite{ADS}. However, its contribution in the sum rule analysis is
tiny and can be quantified by the value of the tachyonic gluon mass  $\lambda^2$
from 
$e^+e^-\to hadrons$ data \cite{CNZ,SNI}, $\pi$-Laplace sum rule \cite{CNZ},  lattice data for the sum of pseudoscalar $\oplus$ scalar two-point correlators \cite{SNZ0}:
\beq
-(\alpha_s/\pi) \lambda^2 = (7\pm 3)\times 10^{-2} ~{\rm GeV}^2~.
\eeq
\vspace*{-0.35cm}
\section{Spectral function }
\nin
\vspace*{-0.9cm}
\subsubsection*{\b Success and test of the na\"\i ve duality ansatz}
\nin
For the light vector and axial-vector channels, the spectral function Im $\Pi(t)$ can be measured inclusively from  $e^+e^-\to hadrons$ and $\tau\to \nu_\tau+ hadrons$ data. In most cases, where hadron masses need to be predicted, there are no data available. Then, one often uses the minimal duality ansatz parametrization of the spectral function:
\beq
{1\over\pi}{\rm Im \Pi(t)}\simeq  f_R^2M_R^{2n}\delta(t-M_R^2)~\oplus~{\rm Im \Pi(t)_{QCD}}~\theta(t-t_c)~,
 \label{eq:ansatz}
\eeq
where: $f_R$ is the resonance coupling to the current;  $n=1,2,..$ is related to the dimension of the current; $t_c$ is the QCD continuum threshold. The QCD continuum comes from the discontinuity of the QCD diagrams in the OPE in order to ensure the matching of the two sides of the sum rules at high-energies. 
The na\"\i ve duality anstaz in Eq. (\ref{eq:ansatz}) works quite well in the different applications of the SVZ sum rules 
within the corresponding expected (10-20)\% accuracy of the approach at LO of PT. A satisfactory test of this model from $e^+e^-\to hadrons$ and charmonium data have been done in \cite{SNB1}. 
\vspace*{-0.25cm}
\subsubsection*{\b Duality violation}
\nin
If one uses the QCD expression of the correlator in the time-like region, the spectral function
may oscillate\,\cite{SHIF,SHIFMAN,RAFA} and can slightly affect (at the \% level\,\cite{PERIS,ALONSO,SNTAU}) the result obtained from the sum rules.  
\vspace*{-0.35cm}
\section{Optimization and Stability Criteria}
\vspace*{-0.1cm}
\nin
There are different steps for improving 
SVZ sum rules:
\vspace*{-0.25cm}
\subsubsection*{\b $\tau$-stability from quantum mechanics}
\begin{figure}[hbt]
\begin{center}
\includegraphics[width=6.5cm]{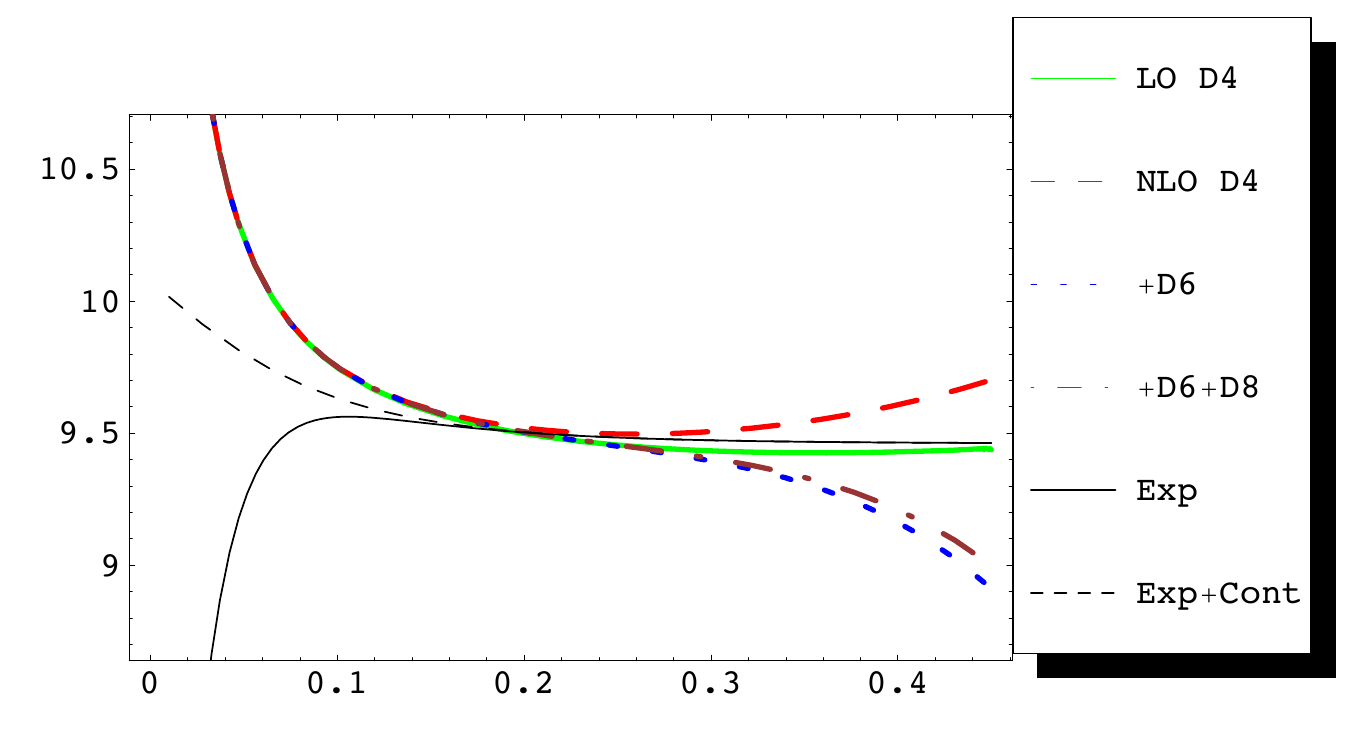}
\caption{\footnotesize  Behaviour of the ratio of moments $\sqrt{{\cal R}^b_0}(\overline{m}_b^2)$ in GeV versus $\tau$ in GeV$^{-2}$ and for $\overline{m}_b(\overline{m}_b) = 4212$ MeV. The black continuous (rep. short dashed) curves are the experimental contribution including (resp. without) the QCD continuum (it is about the $M_{\Upsilon}$). The green (thick continuous) is the PT contribution including the $D=4$ condensate to LO. The long dashed (red) curve is the contribution including the $\alpha_s$ correction to the $D=4$ contribution. The short dashed (blue) curve is the QCD expression including the $D=6$ condensate and the dot-dashed (red-wine) is the QCD expression including the $D=8$ contribution. } 
\label{fig:ratiob}
\end{center}
\vspace*{-0.25cm}
\end{figure}

\nin
\vspace*{-0.45cm}
\begin{center}
\vspace*{-0.5cm}
{\begin{figure}[hbt]
{\includegraphics[width=5.5cm]{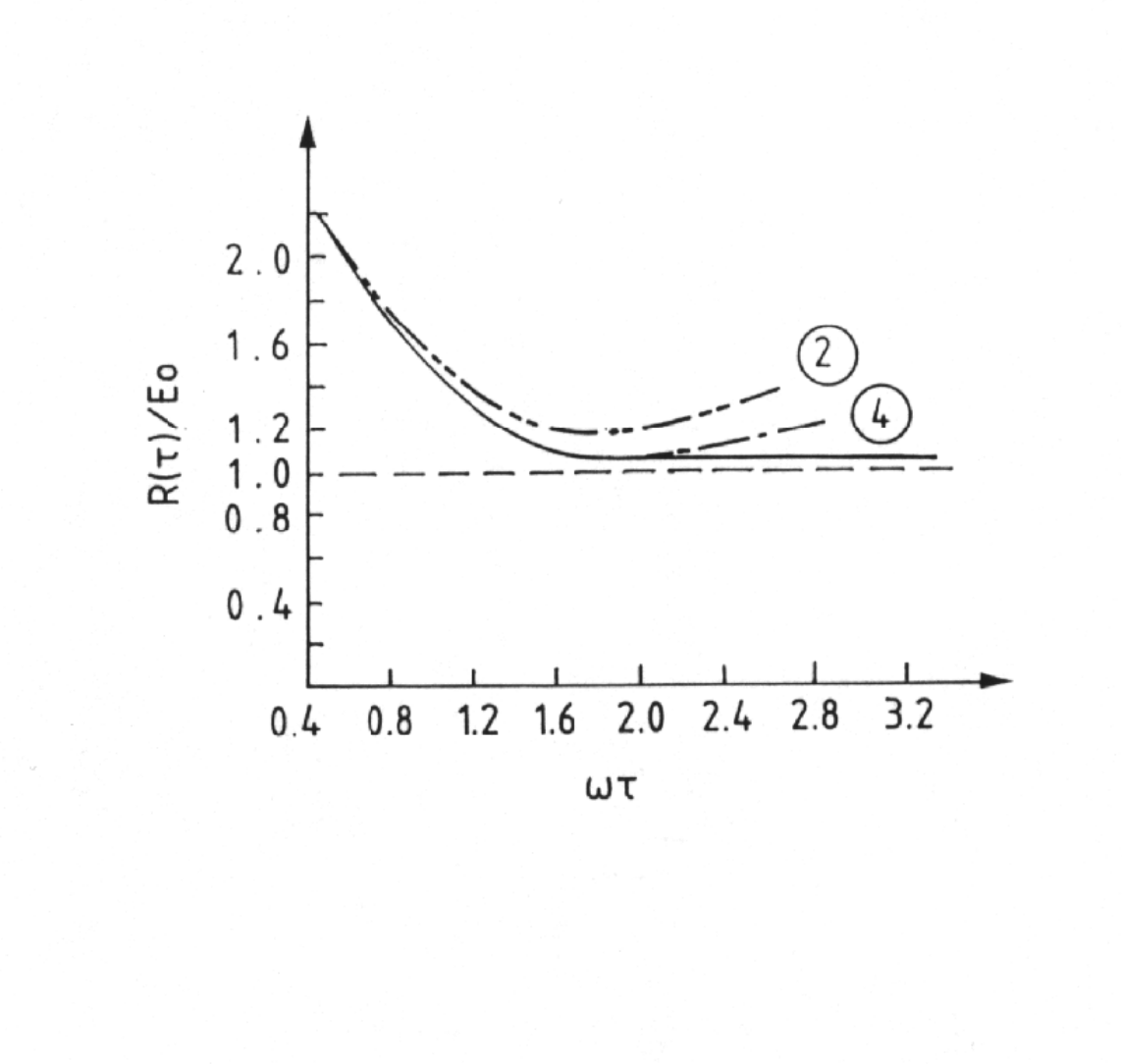}}
\vspace*{-1.5cm}
\caption{Ratio of moments ${\cal R}(\tau)$ in Eq. (\ref{eq:ratio}) versus the sum rule variable $\tau$ (imaginary time) for different truncation of the harmonic oscillator series.}
\label{fig:oscillator}
\vspace*{-0.5cm}
\end{figure}}
\end{center}
\vspace*{-0.5cm}
The question to ask is how one can approach the resonance mass for extracting an optimal information from the sum rules and in the same time the OPE remains convergent. The original SVZ proposal is to find a window where the resonance contribution is bigger than the QCD continuum one and where the non-perurbative terms remain reasonnable corrections. Numerically, the argument is handwaving as the percent of contribution to be fixed is arbitrary. 
In a series of papers, Bell and Bertlmann \cite{BERTL,BELL} have investigated this problem using the harmonic oscillator within the exponential moment sum rules. 
The analysis of the ratio of moments ${\cal R}(\tau)$ defined in Eq. (\ref{eq:ratio}) is
shown in Fig. \ref{fig:oscillator}, where one can observe that the exact solution (ground state energy $E_0$) is reached when more and more terms of the series are added and the optimal information is reached at the minimum 
of ${\cal R}(\tau)$ for a truncated series. We show in Fig. \ref{fig:ratiob}, the $\tau$-behaviour of ${\cal R}(\tau)$, in the bottomium channel, known to N2LO of PT QCD versus the sum rule variable $\tau$ for different truncation of the OPE. 
The position of this minimum co\"\i nc\i de with the SVZ sum rule window but now it becomes more rigorous. 
\nin
\vspace*{-0.25cm}
\begin{center}
\vspace*{-0.5cm}
{\begin{figure}[hbt]
\begin{center}
{\includegraphics[width=6.cm]{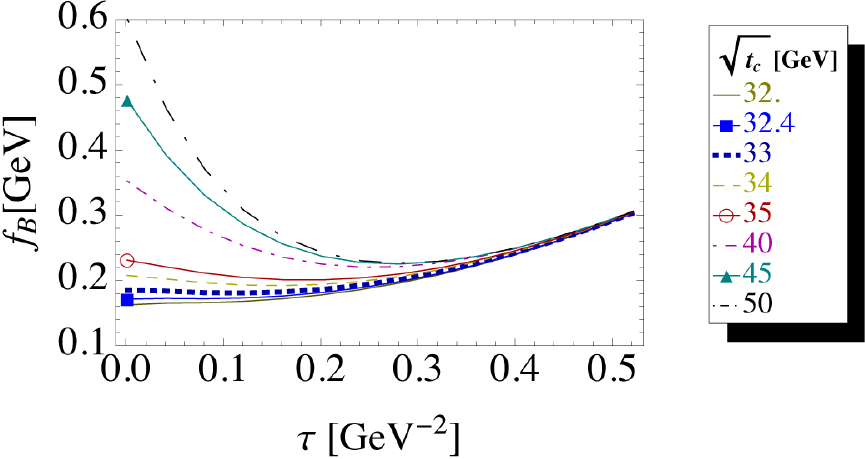}}
\end{center}
\caption{Behaviour of $f_B$ from Eq. (\ref{eq:lapl}) versus the sum rule variable $\tau$ for different values of the QCD continuum threshold $t_c$ and for a given $\mu=$ 3 GeV and $\overline{m}_b(\overline{m}_b) = 4177$ MeV.}
\label{fig:fb}
\vspace*{-0.5cm}
\end{figure}}
\end{center}
\vspace*{-0.25cm}
\vspace*{-0.25cm}
\subsubsection*{\b $t_c$-stability}
\nin
Another free parameter in the phenomenological analysis is the value of the continuum threshold $t_c$. Many authors {\it adjust its value} at the intuitive mass of the next radial hadron excitation. This procedure can be false as the QCD continuum only smears all high mass radial excitations, and what is important is the area in the sum rule integral. As the value of $t_c$ is an external parameter like the $\tau$ sum rule variable, one can also require that the 
physical observables (the lowest resonance parameters) are insensitive to its change. In a such case, we consider the conservative value of $t_c$ from which one starts to have a $\tau$-stability 
until the one where one reaches a $t_c$ stability (see Fig. \ref{fig:fb} for the case of $f_B$). In some cases, this $t_c$ stability is not reached due to the na\"\i ve form of the ansatz in Eq. (\ref{eq:ansatz}). In this case, the complementary use of FESR can help due to the correlation between $t_c$ and the lowest resonance mass (but not with the one of the radial excitation!)\,\cite{PEROTTET}. 
\begin{center}
\vspace*{-0.25cm}
{\begin{figure}[hbt]
\begin{center}
{\includegraphics[width=5.5cm]{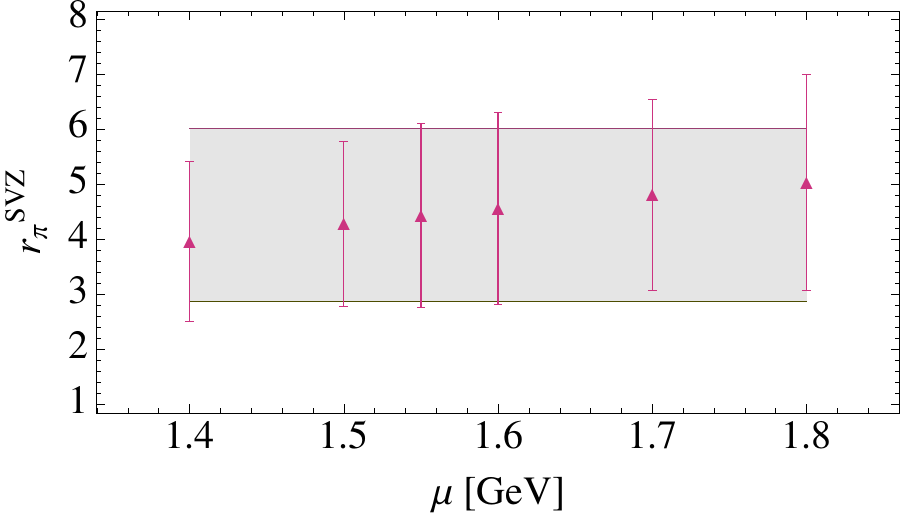}}
\end{center}
\caption{Behaviour of  $r_\pi=(M_\pi'/m_\pi)^4(f_\pi'/f_\pi)^2$ versus  $\mu$. The dashed region  is the mean value with the error from the most precise value.}
\label{fig:rpi}
\vspace*{-0.5cm}
\end{figure}}
\end{center}
\vspace*{-0.25cm}
\subsubsection*{\b New $\mu$-stability criterion}
\nin
Another free parameter in any PT QCD analysis is the subtraction point $\mu$ at which the PT series are evaluated. Like in the previous cases, we consider a minimum sensitivity of the results versus its variation. This is signaled either by a stability plateau, an inflexion point or some extremas\,\footnote{This approach could be an alternative to the optimization of PT series proposed in the current literature \cite{BRODSKY}. We plan to come back to this point in a future work.}. Some illustrative examples based on this criterion have been studied recently in series of papers \cite{SNL14,SNFB12,SNFBST14,SNMOL14}. We show in Fig. \ref{fig:rpi} the $\mu$ behaviour of the ratio $r_\pi=(M_\pi'/m_\pi)^4(f_\pi'/f_\pi)^2$ used to determine  the sum of light quark renomalization group invariant masses: $\hat m_{ud}\equiv(1/2)(\hat m_u +\hat m_d)=\overline{m}_{ud}(-\log{\sqrt{\tau}\Lambda})^{2/-\beta_1}$ shown in Fig. \ref{fig:mud}, where one can notice that the mean value $\la r_\pi\ra=(4.42\pm 1.56)$ from different values of $\mu$ corresponds to the inflexion point at $\mu\simeq 1.55$ GeV \cite{SNL14}; we show in Fig. \ref{fig:fdmu} the $\mu$-behaviour of $f_D$ where a plateau in $\mu$ is obtained \cite{SNFB12} while the ones of the ratio $f_{B^*}/f_B$ and of $f_{B_c}$ are shown in Figs. \ref{fig:rbst} and \ref{fig:fbc} where, here, one has respectively a minimum and an inflexion point \cite{SNFBST14}. 
\begin{center}
{\begin{figure}[hbt]
\begin{center}
{\includegraphics[width=5.5cm]{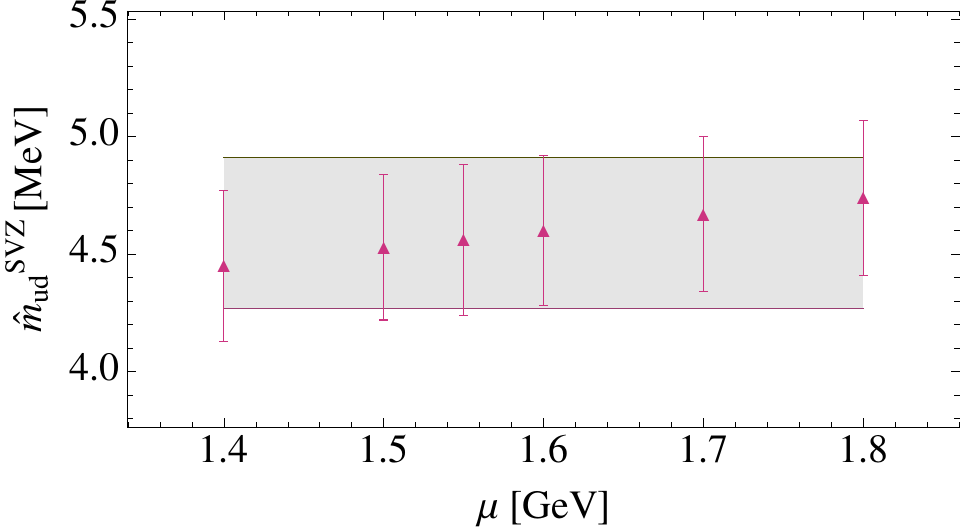}}
\end{center}
\caption{Sum of light quark invariant masses: $\hat m_{ud}$ versus $\mu$. The dashed region is similar as in Fig. \ref{fig:rpi}.}
\label{fig:mud}
\vspace*{-0.5cm}
\end{figure}}
\end{center}
\begin{center}
{\begin{figure}[hbt]
\begin{center}
{\includegraphics[width=4.5cm]{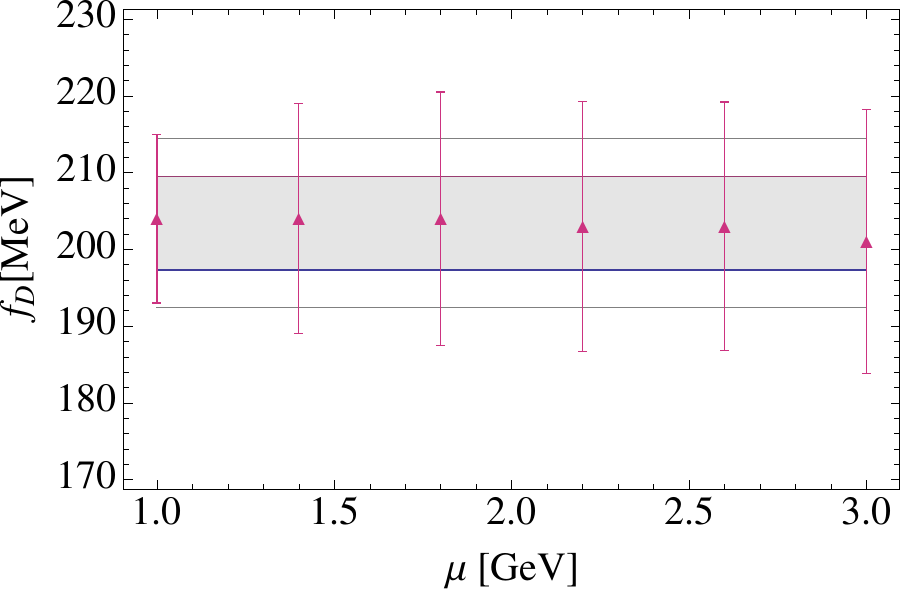}}
\end{center}
\caption{\footnotesize $f_{D}$ versus  $\mu$. The dashed region  is the mean value with the weighted error  while the two upper (lower) lines are when the error comes from the most precise one.}
\label{fig:fdmu}
\vspace*{-0.25cm}
\end{figure}}
\end{center}
\vspace*{-0.25cm}
\begin{center}
\vspace*{-0.25cm}
{\begin{figure}[hbt]
\begin{center}
{\includegraphics[width=4.5cm]{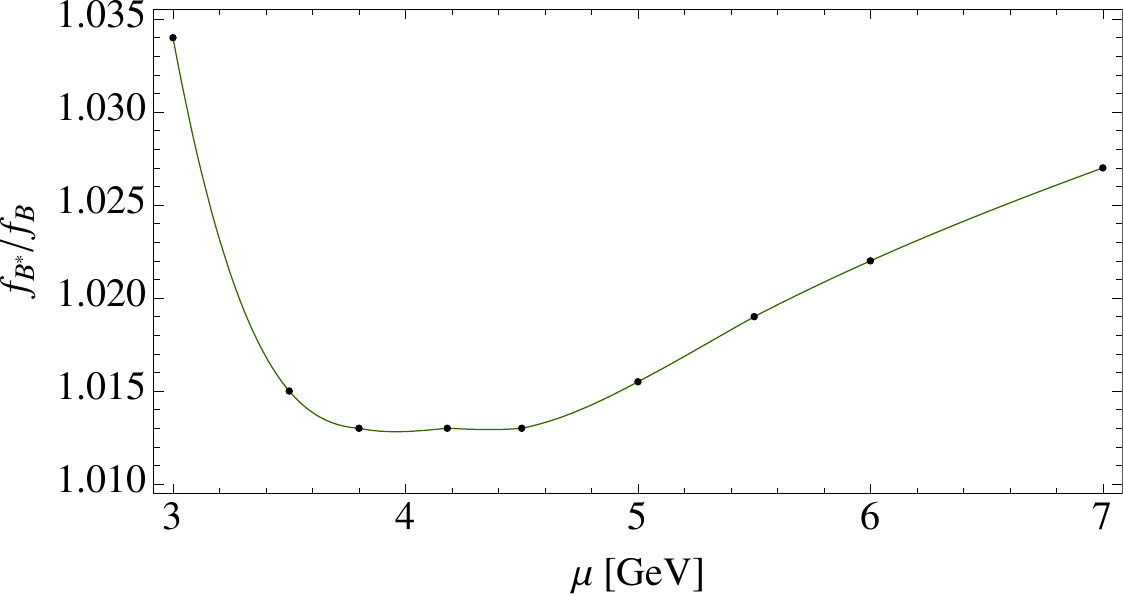}}
\end{center}
\caption{$f_{B^*}/f_B$ versus  $\mu$.}
\label{fig:rbst}
\vspace*{-0.5cm}
\end{figure}}
\end{center}
\begin{center}
\vspace*{-0.25cm}
{\begin{figure}[H]
\begin{center}
{\includegraphics[width=4.cm]{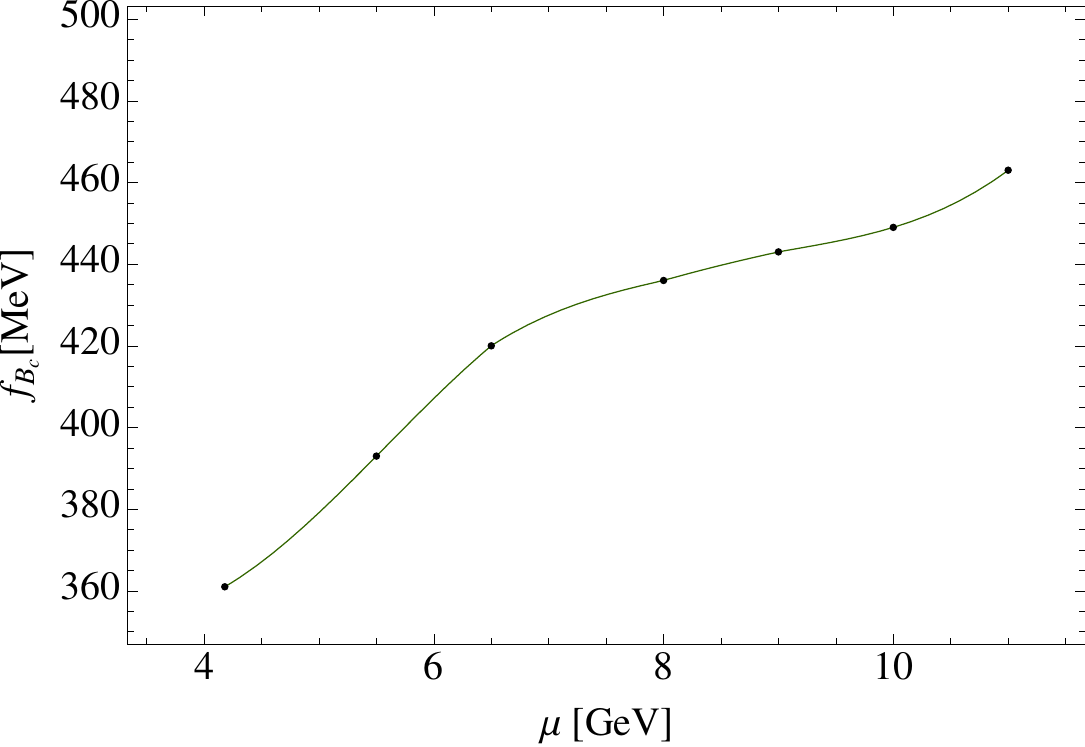}}
\end{center}
\caption{$f_{B_c}$ versus  $\mu$.}
\label{fig:fbc}
\end{figure}}
\end{center}
{\scriptsize
\begin{center}
\begin{table}[hbt]
\setlength{\tabcolsep}{0.15pc}
\begin{tabular}{lllll}
\hline 
\footnotesize  Parameters &\footnotesize $d$&\footnotesize Values [MeV]$^{   d}$&\footnotesize Bounds [MeV]&\footnotesize Sources\\
\hline
\footnotesize   $\alpha_s(M_\tau)$&\footnotesize 0&\footnotesize$0.3249(80)$&&\footnotesize   $\tau$-decay \cite{SNTAU}
\\
\footnotesize$(\alpha_s/\pi) \lambda^2$&\footnotesize 2&\footnotesize$-(265\pm 57)^2$ &&\footnotesize  $\pi$\,\cite{CNZ}\,$\oplus\, e^+e^-$\,\cite{SNI}\,$\oplus\,$ \\
&&&&\footnotesize $\tau$-decay \cite{SNTAU} \\
\footnotesize$\overline{m}_{ud}(2)$&\footnotesize 1&\footnotesize$3.95\pm 0.28$&\footnotesize$\geq 3.28\pm 0.35$&\footnotesize  $\pi$ \cite{SNL14}\\
\footnotesize$\overline{m}_u(2)$&\footnotesize 1&\footnotesize$2.64\pm 0.28$&\footnotesize$\geq 2.19\pm 0.27$&\footnotesize  $\pi\,\oplus\,$ChPT \cite{SNL14}\\
\footnotesize$\overline{m}_d(2)$&\footnotesize 1&\footnotesize$ 5.27\pm 0.49$&\footnotesize$\geq 4.37\pm 0.54$&\footnotesize  $\pi\,\oplus\,$ChPT \cite{SNL14}\\
\footnotesize$\overline{m}_s(2)$&\footnotesize 1&\footnotesize$99.0\pm 5.5$&\footnotesize$\geq 81.6\pm 4.5$&\footnotesize  $K$\,\cite{SNL14}\,$\oplus\, e^+e^-$\,\cite{SNms}\,$\oplus\,$\\
&&&&\footnotesize $\tau$-decay \cite{SNmstau} \\
\footnotesize$\overline{m}_c(\overline{m}_c)$&\footnotesize 1&\footnotesize$1261(12)$&&\footnotesize  $J/\psi,~ D$\,\cite{SNH12,SNFB12,SNH10}\\
\footnotesize$\overline{m}_b(\overline{m}_b)$&\footnotesize 1&\footnotesize$4177(11)$&&\footnotesize  $\Upsilon,~B$\,\cite{SNH12,SNFB12,SNH10}\\
\hline
\end{tabular}
 \caption{\scriptsize  Values of the PT QCD parameters from QSSR: $\overline{m}_{ud}\equiv(1/2)(\overline{m}_u+\overline{m}_d)/2$. The light running masses are evaluated at 2 GeV. }
  \label{tab:pert}
\end{table}
\end{center}
 }
\vspace*{-0.5cm}
\nin
{\scriptsize
\begin{center}
\begin{table}[hbt]
\setlength{\tabcolsep}{0.8pc}
\begin{tabular}{lllc}
\hline 
\footnotesize  Channel  &\footnotesize Values [MeV]&\footnotesize Bounds [MeV]&\footnotesize Sources\\
\hline
\footnotesize $D$ &\footnotesize 204(6)&\footnotesize$\leq 218(2)$& \footnotesize\cite{SNFB12} \\
\footnotesize ${D_s}$ &\footnotesize 246(6)&\footnotesize$\leq 254(2)$& -- \\
\footnotesize $B$ &\footnotesize 206(7)&\footnotesize$\leq 235(4)$& -- \\
\footnotesize ${B_s}$ &\footnotesize 234(5)&\footnotesize$\leq 251(6)$& -- \\
\footnotesize ${D^*}$ &\footnotesize 250(11)&\footnotesize$\leq 266(8)$&\footnotesize\cite{SNFBST14} \\
\footnotesize ${D^*_s}$ &\footnotesize 270(19)&\footnotesize$\leq 287(18)$&-- \\
\footnotesize ${B^*}$ &\footnotesize 209(8)&\footnotesize$\leq 295(15)$&-- \\
\footnotesize ${B^*_s}$ &\footnotesize 225(10)&\footnotesize$\leq 317(17)$&-- \\
\footnotesize ${B_c}$ &\footnotesize 436(40)&\footnotesize$\leq 466(16)$&-- \\
\footnotesize ${D^*_0}$ &\footnotesize 217(24)&& \footnotesize\cite{SNFB04S} \\
\footnotesize ${D^*_{s0}}$ &\footnotesize 202(22)&& -- \\

\hline
\end{tabular}
 \caption{\scriptsize  Heavy-light decay constants $f_H$ within $\mu$-stability at N2LO. }
 \label{tab:decay}
\end{table}
\end{center}
 }
\vspace*{-1.25cm}
\nin
\section{Phenomenological Results}
\nin
QSSR phenomenology is vast and is discussed in details in the previous longer reviews and books \cite{SVZ,SHIF,ZAK,DOSCH,SN10A,ZAKA,BERTL,IOFFE,REV,SNB1,REVMOL}. The common features emerging from these applications are the incredible success of the approach for predicting hadron properties using limited numbers of the universal and fundamental QCD parameters given in Tables \ref{tab:param} and \ref{tab:pert}. This success has been reinforced by the precise extraction of $\alpha_s$ from $\tau$-decay data \cite{BNP,SNTAU,BOITO}.
\vspace*{-0.25cm}
\subsubsection*{\b QCD parameters}
\nin
In addition to the results for the non-perturbative parameters in Table \ref{tab:param}, we show in Table \ref{tab:pert} the values of the PT running QCD parameters obtained from $\tau$-decays and QSSR. The value of $\alpha_s$ is in agreement with the world average and has first shows the running of $\alpha_s$ from $M_\tau$ to $M_Z$. The mass results are improvements of existing ones \cite{SNB1} and,
in general, agree with some other determinations \cite{PDG,FLAG}, while the optimal lower bounds are new.  The r\^ole of the $\pi'(1300)$ and $K'(1460)$ are essential for extracting the light quark masses from the pseudoscalar channels. However, one can notice {\it (en passant)} that the $SU(3)$ breaking ratio of condensates $\kappa\equiv \la \bar ss\ra/\la \bar dd\ra= 0.74(3)$ from different channels (pseudoscalar \cite{SNB1}, light \cite{DOSCH} and heavy\cite{HBARYON}  baryons spectra) and from various authors differs  from recent lattice calculations $\kappa=1.08(16)$ \cite{UKQCD} which needs to be checked. 
\vspace*{-0.67cm}
\subsubsection*{\b Heavy-light decay constants}
\nin
We show in Table \ref{tab:decay} recent values and upper bounds of the heavy light decay constants obtained in \cite{SNFB12,SNFBST14,SNFB04S}. These results are in good agreement with older ones \cite{SNB1,SNFB}, with lattice calculations \cite{ROSNER,BECIR} and other recent QSSR estimates  \cite{ROSNER,PIVOV}.  
\vspace*{-0.25cm}
\subsubsection*{\b Exotic  mesons}
\nin
Exotics, especially charmonium and bottomium states have been intensively discussed these last few years (see e.g. \cite{CONF,REVMOL,MATHEUS} for molecules and four-quark states; see e.g. \cite{STEELE} for heavy and \cite{SNHYB} for light hybrid states). However, in most  works for molecules and four-quark states, it is disappointing that the two-point correlator is only known at LO of PT QCD where the  heavy quark masses used  are ill-defined while some classes of non-perturbative condensates are known until dimension $(10 - 12)$ but 
some others like e.g. the gluon condensates contributions are not included in the OPE. These weak points have been cured in recent works where N2LO PT corrections have been included \cite{SNMOL14} which render more reliable the QSSR predictions. The observed $Z_c(3900)$ and $Z_b(10610)$ almost co\"\i ncides with the predictions 3738(152)\,MeV and 10687(232)\,MeV for \,$\bar D^*D$ and $\bar B^*B$\,$(1^{++})$ molecules, while,  
the experimental candidates $Y_c(4260),~Y_c(4360)$ and $Y_c(4660)$ are too light compared with the prediction of $(5176\pm 224)$\,MeV for  a pure $\bar D^*D^*_0$\,$(1^{--})$ molecular state.
\vspace*{-0.25cm}
\subsubsection*{\b Glueball / gluonium and nature of $\sigma$ meson}
\nin
Another recent QSSR application is the eventual possibility for the $\sigma$-meson to be the lowest
scalar gluonium while its large coupling to $\pi^+\pi^-$ and to $K^+K^-$ can be due to the large violation of the OZI rule
at this low energy. This result has been deduced by comparing QSSR predictions for a glueball mass around 1 GeV \cite{VENEZIA} needed for a self-consistency of the (un)subtracted sum rules (similar result is found from lattice  and a NJL model \cite{FRASCA})  with the phenomenological analysis of $\pi\pi$ and $\gamma\gamma$ scattering data using K-matrix approach which can separate the pole from the rescattering contribution \cite{MENES}. This feature of the $\sigma$ has to be contrasted to the favoured G(1.6) glueball candidate where at this energy one expects a much better realization of the OZI rule such that it is expected to decay into pairs of $\eta'\eta,~\eta\eta~U(1)$ gluonic states but couples weakly to Goldstone pairs \cite{VENEZIA}. 
\vspace*{-0.25cm}
\section*{Conclusions}
\vspace*{-0.1cm}
\nin
We have presented in a compact form the developments of QCD spectral sum rules 35 years later. 
Values of the QCD parameters and some decay constants from QSSR are summarized in Table \ref{tab:param} to \ref{tab:decay}. Its applications are rich and have the advantage (compared to lattice simulations) to be analytical where one can control the size of different contributions. The successful applications of the sum rules motivate further studies and improvements of the approach by e.g. incuding radiative and quadratic corrections.
\vspace*{-0.4cm}

\end{document}